\documentclass[12pt]{article}
\usepackage{amsmath}
\usepackage{amsfonts}
\usepackage{graphicx}
\usepackage{amssymb}
\usepackage{yhmath}
\usepackage{pdfpages}
\graphicspath{ {images/} }

\newtheorem{Def}{Definition}[section]

\newcommand{\R}{\mathbb{R}}

\def \seq {\subseteq}
\def \qed {\hfill \vrule height6pt width 6pt depth 0pt}
\def \noi {\noindent}

\begin{document}

\title{On spatial majority voting with an even (vis-a-vis odd) number of voters: a note}
\author{Anindya Bhattacharya\thanks {(Corresponding Author) Department of Economics and Related Studies, University of York, UK; Email: anindya.bhattacharya@york.ac.uk;}~~~Francesco Ciardiello\thanks{Department of Economics and Statistics (DISES), University of Salerno, Italy; Email: fciardiello@unisa.it.}}

\date{\today}

\maketitle

\begin{abstract}

In this note we consider situations of (multidimensional) spatial majority voting. We show that under some assumptions usual in this literature, with an {\em even} number of voters if the core of the voting situation is singleton (and in the interior of the policy space) then the element in the core {\em is never a Condorcet winner.} This is in sharp contrast with what happens with an {\em odd} number of voters: in that case, under {\em identical} assumptions, it is well known that if the core of the voting situation is non-empty then the single element in the core is the Condorcet winner as well.
   
\end{abstract}

\vskip3em

\noi {\em  Keywords:} Spatial Voting Situations; Core; Condorcet winner. \\

\noi

\newpage

\section{Introduction}

\noi It is quite well-known that for a voting situation if the number of voters changes from being odd to even then the properties associated with the voting situation, {\em everything else remaining the same}, can change. Perhaps the most well-known example of such changes is the following. Take a finite set of voters and a finite set of candidates and suppose that each voter has a strict preference ordering over the set of candidates. Then, if there are an odd number of voters then, with majority rule voting, for any two candidates $x$ and $y,$ either $x$ is socially preferred to $y$ or the converse is true: i.e., the resulting social ordering generates a strict tournament. But that is no more true in general if there are an even number of voters.\\

\indent In this note we provide another (possibly interesting) example of such a phenomenon in the context of (multidimensional) {\em spatial} voting with majority rule. We show that under some assumptions usual in this literature, with an {\em even} number of voters if the core of the voting situation is singleton (and in the interior of the policy space) then the element in the core {\em is never a Condorcet winner.} This is in sharp contrast with what happens with an {\em odd} number of voters: in that case, under {\em identical} assumptions, it is well known that if the core of the voting situation is non-empty then the single element in the core is the Condorcet winner of the voting situation as well.\\

\indent At first sight our finding may not seem interesting: indeed, one might think that when the binary relation ``a strict majority of voters prefer $x$ to $y$" is not complete then it is no big deal that ``beats all others" and ``is beaten by no other" may not necessarily coincide. In response we would like to point out that our result does not point out {\em one/some ``may be" case(s).} We show, we repeat, that if the number of voters changes from being odd to even then, under some usual regularity assumptions, {\em excatly for half of the possible cases} a property of the core (perhaps, justifiably, the most intensively-studied solution for this environment) gets, somewhat ``dramatically", {\em exactly opposite}!\\  

\indent The next section gives the preliminary definitions and some pieces of notation. The main result and discussions around it are given in Section 3. Section 4 provides some concluding remarks. At the end, via a diagram we illustrate one idea behind the proof of the main proposition.

\section{Preliminary definitions and notation}

\noindent
Let $Z\subseteq \R^{k}$ be a compact full-dimensional convex subset of some finite ($k$-)dimensional Euclidean space with $k \geq 1.$ This set, $Z,$ is identified to be the feasible set of policies or outcomes on which a voter votes. Let $N$ be the finite set of players or voters. Suppose that the preferences of a player $i$ on $Z$ is represented by a real-valued continuously differentiable and strictly concave pay-off function $u_{i}\in C^{1}(Z, \R).$ The spatial voting situation we consider below is obtained by introducing the method of majority rule voting.
\begin{Def}[Domination by Majority Rule]
Given $x,y \in Z,$ the policy $x$ beats (or dominates) policy $y$ via
coalition $S \seq N,$ if and only if $|S|>|N|/2${\footnote{For a set $A,$ by $|A|$ we denote its cardinality.}} and $u_i(x)>u_i(y)$
for each $i \in S.$ We denote this as $x \succ_S y.$ If there exists
a majority coalition $S$ via which $x$ dominates $y,$ we denote that
as $x \succ y.$
\end{Def}
The collection $G=(Z,N,(u_i)_{i \in N})$ is a spatial voting
situation with majority rule.\\

\noi For any $x \in Z$ and $i \in N,$ by $D^i(x)$ we denote the set $\{y \in Z \,:\, u_i(y) \geq u_i(x)\}.$ For any set $A \seq Z,$ by $cl(A)$ and $int(A)$ we denote the closure of $A$ and the interior of $A$ respectively. Also, for any two points $x,y \in Z,$ by $\rho(x,y)$ we denote the (Euclidean) distance between these two points.\\

\noindent
Recall the notion of the core for such a situation.

\begin{Def}[The Core of a Voting Situation]
The core of such a voting situation $G$ is the subset
$
K=\{y\in Z \,:\,\nexists x \in Z \,\,\mbox{such that}\,\, x \succ y \}.
$
\end{Def}

\noi Recall that a point $x \in Z$ is said to be the Condorcet winner of the voting situation if for any other outcome $y \neq x,$ $x \succ y.$ Recall that if a voting situation admits a Condorcet winner, then it is the unique element in the core. 

For $i \in N,$ we denote the unique maximizer of $u_i$ on $Z$ by $\bar{x}_i.$ Similarly, for any (one-dimensional) straight line $L$ and any convex subset $A \seq Z,$ such that $L \cap A$ is non-empty, we denote the unique maximizer of $u_i$ on the convex set $L \cap A$ by $\bar{x}_i (L \cap A)$ and call it the {\em induced ideal point of voter $i$ on the line segment $L \cap A.$}

\section{The result and some discussions}

\noi The main result is:\\

\noi {\bf{Proposition 1}} {\em{Consider a voting situation $G$ for which $|N|$ is an even (say, $2n$ where $n \geq 1$) positive integer. Suppose further that for $G,$ the core $K=\{z\}$ is singleton, $z$ is in the interior of $Z,$ and for no more than one $i \in N$ is it the case that $z=\bar{x}_i.$ Then $z$ cannot be the Condorcet winner for $G.$}}\\

\noi {\em Remark 1} Note that if $n=1,$ i.e., there are only two voters $(i,j),$ then the core is singleton if and only if the element in the core $z=\bar{x}_i = \bar{x}_j.$\\

\noi Recall the property of the core under identical assumptions when $|N|$ is odd:\\

\noi {\bf{Proposition $1^{\prime}$}} {\em{Consider a voting situation $G$ for which $|N|$ is an odd positive integer. Suppose further that for $G$ the core $K$ is non-empty. Then there is a unique element in the core, $z,$ which is the Condorcet winner for $G.$}}\\

\noi For proof of Proposition $1^{\prime}$ see, if necessary, Cox (1987, p. 411) and Exercise 5.5 in Austen-Smith and Banks (1999).\\

\noi Before proceeding to the proof of Proposition 1 we collect here a couple of useful facts and an elementary lemma.\\

\noi First recall that for any (one-dimensional) straight line $L \subset \R^{k},$ and a point $y \in L,$ one can decompose $L$ into two half-lines $H^+_y(L)$ and $H^-_y(L)$ such that $H^+_y(L) \cap H^-_y(L)=\{y\}$  (see, if necessary, p. 135 of Austen-Smith and Banks, 1999). Recall further the following.\\

\noi {\bf Fact 1} Let $L$ be any (one-dimensional) straight line ($L \subset \R^{k}$). Suppose further that for some convex $A \seq Z,$ $L \cap A \neq \emptyset.$ Then, for each $i \in N,$ $i$'s preference restricted to $L \cap A$ is single-peaked with $\bar{x}_i (L \cap A),$ the unique maximizer of $u_i$ on the convex set $L \cap A,$  being the peak.\\

\noi For the proof, again, see, if necessary, p. 135 of Austen-Smith and Banks (1999).\\

\noi {\bf{Lemma 1}} {\em{Suppose $z$ belonging to the interior of $Z$ is the single point in the core and that it is the Condorcet winner as well. Then for any straight line $L$ with $(\{z\} \cap L) \neq \emptyset,$ there exists at least a pair of voters $(i,j)$ such that for $m \in \{i,j\},$ $\bar{x}_m (L \cap Z),$ the unique maximizer of $u_m$ on the convex set $L \cap Z,$ is $z.$}}\\

\noi {\bf Proof of Lemma 1.} Suppose not. Then, since $z$ is in the core, each of the half-lines $H^+_z(L)$ and $H^-_z(L)$ must contain exactly $n$ (i.e., $|N|/2$) of the induced ideal points on $Z \cap L$ (as, otherwise, by Fact 1, $z$ would be dominated). Consider $i$ and $j$ such that\\ 
$\bar{x}_i (L \cap Z) \in  H^+_z(L)$ and for every $l \in N$ for whom $\bar{x}_l (L \cap Z) \in  H^+_z(L),$ $\rho(\bar{x}_l (L \cap Z),z) \geq \rho(\bar{x}_i (L \cap Z),z);$ and\\
$\bar{x}_j (L \cap Z) \in  H^-_z(L)$ and for every $l \in N$ for whom $\bar{x}_l (L \cap Z) \in  H^-_z(L),$ $\rho(\bar{x}_l (L \cap Z),z) \geq \rho(\bar{x}_j (L \cap Z),z).$\\
Then, by Fact 1 above (i.e., for every voter $m,$ the induced preference on $L$ is single-peaked around $\bar{x}_m (L \cap Z)$) $z$ cannot dominate any of the points lying on the line-segment $(\bar{x}_i (L \cap Z), \bar{x}_j (L \cap Z))$ contradicting the supposition that $z$ is the Condorcet winner. \qed\\

\noi {\bf Proof of Proposition 1.} First we consider a voting situation $G$ where $k=1:$ i.e., $Z$ is a either a single point or a straight line segment. Then, by Lemma 1, if $z$ is the Condorcet winner of $G$ then there exists at least a pair of voters $(i,j)$ such that $z=\bar{x}_i=\bar{x}_j.$ But that violates the assumption in the Proposition that for no more than one $i \in N$ is it the case that $z=\bar{x}_i.$\\

\noi Next we consider the case where $k>1.$ Suppose a voting situation $G$ obeys the assumptions in Proposition 1 but for which $z,$ the singleton point in the core, is the Condorcet winner of $G.$\\ 

\noi Now call the voter $l$ (if such a voter exists) for whom ${\bar {x}}_l=z.$ Note that for any voter ${i} \in N\setminus \{l\},$ given our assumptions on $u_i,$ $D^i(z),$ the upper contour set for ${i}$ at $z,$ is a convex subset (of dimension $k$) of $Z.$ Further, again for any $i \in N \setminus \{l\},$ given our assumptions on $u_i,$ by Theorem 3.1 of He and Xu (2013) $D^i(z)$ has a unique supporting hyperplane at $z,$ one of its boundary points.\\ 

\noi Take a closed ball $B \seq Z$ with $z$ as the centre of $B$ (which is assured as $z$ is in the interior of $Z$). 

For any point $x$ in the boundary of $B$ call the set of points in $B$ which belong to the straight line connecting $x$ and $z$ a diameter of $B$ (we specify this because ``diameter'' of a set has been used in literature to define a different concept also). Then

$$B={\displaystyle \cup} \{L \mid {\mbox{$L$ is a diameter of $B$ (passing through $z$)\}.}}$$

\noi Since $z$ is assumed to be the Condorcet winner, by Lemma 1 above, for each such diameter $L,$ there exists a pair of voters $(i(L),j(L))$ such that for $m \in \{i(L), j(L)\},$ $\bar{x}_{m} (L \cap B),$ the unique maximizer of $u_{m}$ on the convex set $L \cap B,$ is $z.$\\
However, since $N$ is a finite set, there are only finitely many such possible pairs. Therefore, there must exist a pair of voters $(\bar{i}, \bar{j})$ such that for a convex subset $\Delta \seq B,$ (of dimension $k$) of diameters, the following has to be true:\\
(i) for each diameter $L' \subset \Delta,$ $\bar{x}_{\bar{i}} (L' \cap B)=z;$\\
(ii) for at least one diameter $L,$ $L \cap int(D^{\bar i}(z)) \neq \emptyset$ (as there exists only {\em one} supporting hyperplane to $D^{\bar i}(z)$ at $z$).\\
Note that such a voter $\bar{i}$ exists even if the other voter in the pair is $l:$ i.e., for whom ${\bar {x}}_l=z.$\\ 

\begin{center}
[Insert Figure 1 here. Figure caption: ``A geometrical illustration (for two dimensions) of the argument for proving Proposition 1".]
\end{center}

\noi But (i) and (ii) above cannot be true together. \qed\\

\noi {\em Remark 2} Note that if $|N|$ is odd and the the element in the core $z$ is in the interior of $Z,$ then at least for one voter $l,$ ${\bar {x}}_l=z$ (see, if necessary, Austen-Smith and Banks 1999, p. 143). So, the parallelism between such a situation with $|N|$ being odd and the set-up for Proposition 1 is quite clear. Also, note that the condition in Proposition 1 that $z,$ the unique element in the core is in the interior of $Z,$ and that for no more than one $i \in N$ is it the case that $z=\bar{x}_i$ is generic.

\section{Concluding remarks}

\noi Here we indicate one implication of Proposition 1. Bhattacharya et al. (2018) looked at the relation between the (Gillies) uncovered set and the core under identical assumptions as in this paper. It is well-known that under these assumptions, when there are an odd number of voters, if the core is non-empty then the singleton core is the (Gillies) uncovered set as well (this result goes back, at least, to Cox 1987). But Bhattacharya et al. (2018) showed that if the number of voters is even then a singleton core is the uncovered set if and only if the unique element in the core is the Condorcet winner as well (and thus, with an even number of voters, the uncovered set may not coincide with the core). Therefore, as a corollary of Proposition 1 we obtain that      
if the number of voters is even then, under the regularity assumptions of the Proposition, a singleton core is {\em never} the (Gillies) uncovered set which, again, is in sharp contrast with what happens with an {\em odd} number of voters.

\begin{center}
\begin{figure}[htbp]
\begin{center}
\includegraphics[angle=0, scale=0.60]{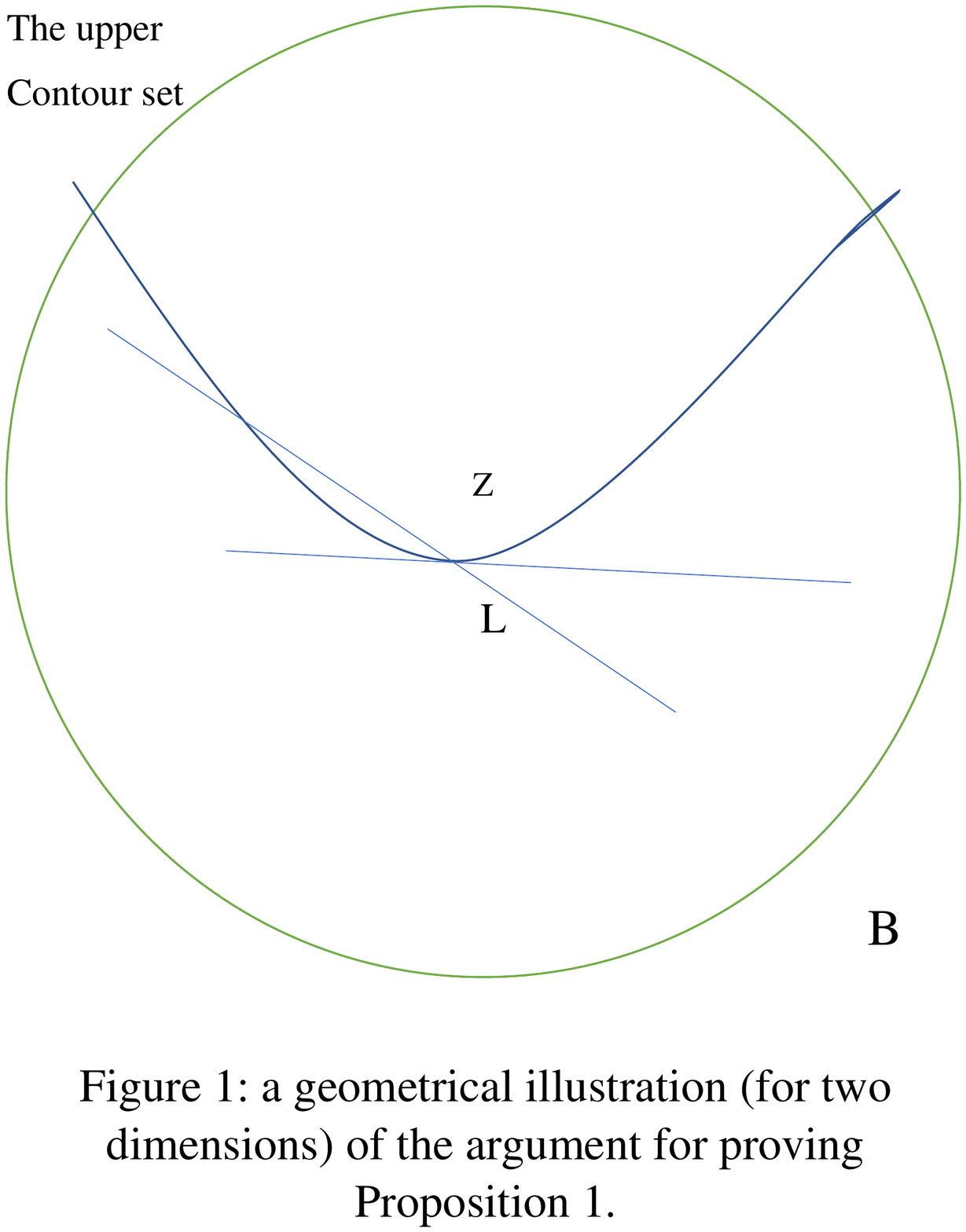}
\label{default}
\end{center}
\end{figure}
\end{center}

\end{document}